\documentclass[%
 iop,superscriptaddress,
 jcp,%
 amsmath,amssymb,floatfix,
 reprint,twocolumns%
]{revtex4-1}

\usepackage[version=3]{mhchem} %

\usepackage{dcolumn}%
\usepackage{bm}%
\usepackage{todonotes}
\usepackage{tikz}
\usepackage{nicefrac}
\usepackage{array}
\usepackage{amssymb}
\usepackage{amsmath}
\usepackage{graphicx}
\usepackage{multirow}
\usepackage{color}
\usepackage{comment}
\usepackage[normalem]{ulem}
\usepackage{url}
\usepackage{mathrsfs}
\usepackage{subfigure}
\usepackage{float}

\newcommand{\D}{\ensuremath{\mathrm{d}}}

\newcommand\arS{\rule{0pt}{12pt}}

\begin{document}

\title{Fine Tuning Classical 
and Quantum Molecular Dynamics\\ using a Generalized
Langevin Equation
}%

\author{Mariana Rossi}
\affiliation{Laboratory of Computational Science and Modelling, Institute of Materials, Ecole Polytechnique F\'ed\'erale de Lausanne, Lausanne, Switzerland}
\affiliation{Current Address: Fritz Haber Institute of the Max Planck Society, Faradayweg 4-6, 14195 Berlin, Germany}
 
\author{Venkat Kapil}
 \affiliation{Laboratory of Computational Science and Modelling, Institute of Materials, Ecole Polytechnique F\'ed\'erale de Lausanne, Lausanne, Switzerland}

\author{Michele Ceriotti}
\email{michele.ceriotti@epfl.ch}
\affiliation{Laboratory of Computational Science and Modelling, Institute of Materials, Ecole Polytechnique F\'ed\'erale de Lausanne, Lausanne, Switzerland}%

\date{\today}%

\begin{abstract}

Generalized Langevin Equation (GLE) thermostats have
been used very effectively as a tool to 
manipulate and optimize the sampling of 
thermodynamic ensembles and the associated
static properties. Here we show that a 
similar, exquisite level of control can
be achieved for the dynamical properties 
computed from thermostatted trajectories. 
By developing quantitative measures of the disturbance 
induced by the GLE to the 
Hamiltonian dynamics of a harmonic
oscillator, we show that 
these analytical results accurately predict 
the behavior of strongly anharmonic systems.
We also show that it is possible to correct, to a 
significant extent, the effects of  
the GLE term onto the corresponding microcanonical 
dynamics, which puts on more solid grounds the use of non-equilibrium
Langevin dynamics to approximate quantum nuclear effects and could help
improve the prediction of dynamical quantities from techniques that use a Langevin term to stabilize dynamics.
Finally we address the use of thermostats in the
context of approximate path-integral-based models
of quantum nuclear dynamics. We demonstrate that
a custom-tailored GLE can alleviate some of
the artifacts associated with these techniques, improving
the quality of results for the modelling
of vibrational dynamics of molecules, liquids and solids.
\end{abstract}

\maketitle

\section{Introduction}

Dynamical properties of the ionic degrees of freedom of a material or a molecule 
provide a direct connection to experimental observables such as vibrational spectra, diffusion
coefficients, reaction rates and heat conductance, among others. Their evaluation from atomistic simulations
is typically more challenging than the evaluation of static ensemble properties, especially when taking
into account nuclear quantum effects (NQEs). To name only a few reasons for this greater challenge, one must ensure sufficient sampling, but in a standard simulation thermostats often cannot be used to aid this task, since their presence modifies dynamical properties.
Moreover, no computationally affordable exact method to include nuclear quantum
effects in dynamical properties exists, and the many approximate methods available that are based on path integral
molecular dynamics \cite{cao-voth94jcp, crai-mano04jcp, ross+14jcp}, suffer from one or another unphysical artifact \cite{habe+08jcp,witt+09jcp,ross+14jcp}.

Generalized Langevin Equation (GLE) thermostats emerged as an efficient tool for the control
and evaluation of static properties in both classical and quantum nuclei simulations \cite{ceri+09prl,ceri+09prl2,ceri+10jctc}.
In order to obtain a targeted GLE kernel for the static ensemble properties of interest,
one can define, in a reasonably straightforward manner, target quantities that measure
the performance of the GLE dynamics when applied to the actual system\cite{ceri+10jctc}. By defining different fitting
targets, it is possible to build thermostats that act only in a specific set of modes~\cite{ceri+09prl},
that are efficient in a wide frequency range~\cite{ceri+10jctc},
that mimic nuclear quantum fluctuations~\cite{ceri+09prl2}, and many other possibilities~\cite{ceri-parr10pcs,dett+17jctc}.

In this paper, we study and show how GLE thermostats can be used
in order to control {\it dynamical properties} of various systems.
We develop a framework for the optimization of GLE matrices where 
we construct simple dynamical models and define new target quantities sensitive to dynamical information. 
We take as paradigmatic examples the vibrational spectra of water ranging from the isolated
molecule and protonated clusters, all the way to the condensed phase.
On one hand, we show how the disturbance due
to the GLE dynamics in classical nuclei simulations can be predicted, and how the perturbed spectra can be deconvoluted to recover the unperturbed density of states.
On the other hand, we take advantage of the freedom that thermostatted ring polymer molecular dynamics (TRPMD) \cite{ross+14jcp}
leaves in the choice of the thermostat attached to the internal modes of the ring polymer
in order to design GLE matrices that reduce the spurious broadening of high-frequency spectral features  
that was observed in the original formulation based on a white noise thermostat \cite{ross+14jcp,ross+14jcp2}.
In both cases, we use the same framework to tune the dynamics at will.

Throughout this work, we take also special attention to perform our dynamics using
accurate potential energy surfaces that include all physically relevant effects
for the systems in question. For the molecules, we use extremely accurate parametrized
potentials\cite{part-schw97jcp,HuangBowman2005}, and for the condensed phase simulations we use neural network potentials
fitted to accurate density-functional theory data\cite{morawietz2016,chen+16jpcl}. 
In Section II we explain our models and fitting procedures in detail, showcasing
the rationale of our development with toy examples. In Section III we first detail how we
can predict and deconvolute the GLE disturbance on classical molecular dynamics,
and then show how fitted GLE matrices can change the outcome of TRPMD simulations.
Finally, in Section IV we draw our conclusions.

\section{Theory and Methods}

The use of a history-dependent Langevin equation 
to model the coupling between 
a system and a canonical heat bath 
has been discussed many times~\cite{zwan61pr}. 
Such generalized Langevin equations (GLEs) have
been studied extensively as a tool to study reaction rates~\cite{tuck+91jcp}, to model open
systems~\cite{stel+14prb}, and as a general sampling
device whose properties can be formally quantified~\cite{otto+12jfa,hall+16jcp}.
The use of a GLE as a highly tunable
thermostatting scheme for atomistic simulations 
has also been discussed extensively elsewhere~\cite{ceri10phd,ceri+10jctc}. 
For the sake of completeness and 
to introduce notation, we will 
briefly summarize the basic ideas, 
before discussing in more detail 
how this GLE framework can be used to 
obtain a precise control of 
the dynamics of a physical system.

\subsection{A Generalized Langevin Equation Thermostat}

The generalized Langevin equation for a 
particle with unit mass in one dimension, subject to a 
potential $V(q)$, is given by the 
non-Markovian process
\begin{equation}
\label{eq:nonMarkov}
\begin{split}
\dot{q}&=p\\
\dot{p}&=-V'(q)- \int_{-\infty}^t K(t-s) p(s) \mathrm{d} s +\zeta(t)
\end{split}
\end{equation}
where $K(t)$, is the memory kernel 
that describes dissipation,
and $\zeta(t)$ is a Gaussian random 
process with a time correlation 
function $H(t) = \langle\zeta(t)\zeta(0)\rangle$. 
Throughout this paper, we consider unity mass in all equations. 
The numerical integration of this 
equation is computationally challenging 
since it requires the knowledge of the 
entire history of the particle's 
trajectory. However, exploiting the equivalence between the non-Markovian dynamics of Eq. \ref{eq:nonMarkov} and 
Markovian dynamics in an extended space, $n$ auxiliary degrees of freedom 
$\mathbf{s}$ can be coupled linearly to physical momenta, which results in the Markovian Langevin equation
\begin{equation}
\begin{split}
  \dot{q}=&p\\
\!\left(\! \begin{array}{c}\dot{p}\\ \dot{\mathbf{s}} \end{array}\!\right)\!=&
\left(\!\begin{array}{c}-V'(q)\\ \mathbf{0}\end{array}\!\!\right)
\!-\!\left(\!
\begin{array}{cc}
a_{pp} & \mathbf{a}_p^T \\ 
\bar{\mathbf{a}}_p & \mathbf{A}
\end{array}\!\right)\!
\left(\!\begin{array}{c}p\\ \mathbf{s}\end{array}\!\right)\!+\!
\left(\!
\begin{array}{cc}
b_{pp} & \mathbf{b}_p^T \\ 
\bar{\mathbf{b}}_p & \mathbf{B}
\end{array}\!\right)\!
\left(\!\begin{array}{c}\multirow{2}{*}{$\boldsymbol{\xi}$}\\ \\\end{array}\!\right).
\end{split}
\label{eq:Markov}
\end{equation}
Here $\boldsymbol{\xi}$ is a $n+1$ dimensional vector of uncorrelated Gaussian numbers. 
In order to label the portions of the matrices that describe the coupling 
between the different components of the extended state vector $\mathbf{x} \equiv (q,p,\mathbf{s})^{T}$, 
we use the following notation:
\begin{equation}
\begin{array}{ccccc}
      &   q   &    p   &   \mathbf{s}  & \arS \\ \cline{2-4}
\multicolumn{1}{c|}{q} & m_{qq} & m_{qp} & \multicolumn{1}{c|}{\mathbf{m}_q^T} & \arS \\\cline{3-4}
\multicolumn{1}{c|}{p} & \multicolumn{1}{c|}{\bar{m}_{qp}} &  m_{pp} &  \multicolumn{1}{c|}{\mathbf{m}_p^T} & \arS \\\cline{4-4}
\multicolumn{1}{c|}{\mathbf{s}} &  \multicolumn{1}{c|}{\bar{\mathbf{m}}_q}  &  \multicolumn{1}{c|}{\bar{\mathbf{m}}_p} &  \multicolumn{1}{c|}{\mathbf{M}} & \arS \\\cline{2-4}
\end{array}
\hspace{-8pt}\begin{array}{cc}
\arS \\ \arS \\
\left.\rule{0pt}{12pt}\right\}\!\mathbf{M}_p \\
\end{array}
\hspace{-8pt}\begin{array}{cc}
\arS \\
\left.\rule{0pt}{20pt}\right\}\!\mathbf{M}_{qp} 
\end{array}
\label{eq:notation}
\end{equation}
Upon integrating out the auxiliary degrees of freedom, equation \ref{eq:nonMarkov} is recovered with 
\begin{equation}
\begin{split} 
K(t)=&2a_{pp} \delta(t)-\mathbf{a}_p^T e^{-\left|t\right|\mathbf{A}}\bar{\mathbf{a}}_p\\
H(t)=& d_{pp} \delta(t)-\mathbf{a}_p^T e^{-\left|t\right|\mathbf{A}}\left[\mathbf{Z}\mathbf{a}_p-\mathbf{d}_p\right]
\end{split}
\label{eq:dred-mem-t}
\end{equation}
where $\mathbf{Z}=\int_0^\infty e^{-\mathbf{A}t}\mathbf{D} e^{-\mathbf{A}^T t}\mathrm{d}t$ and 
$\mathbf{D}_{p} = \mathbf{B}_{p}\mathbf{B}_{p}^{T}$. 
This implies that by tuning the elements of the matrices $\mathbf{A}$ and $\mathbf{B}$, a Generalized Langevin equation with the desired friction kernel and noise 
correlation can be approximated within a Markovian framework. 
Note that although we focused on a one-dimensional
case to simplify the notation, it is also possible to apply
Eqn.~\eqref{eq:Markov} to each Cartesian coordinate of an 
atomistic system. Since the overall dynamics is invariant
to a unitary transformation of the coordinates, the response
of the system would be the same as if the GLEs had been applied
in e.g. the normal modes coordinates.

\subsection{Controlling Classical Dynamics}

 Let us consider a particle subject to a harmonic potential $V(q) = \frac{1}{2}\omega_{0}^{2}$, and coupled to
a GLE. The time evolution of its state vector $\mathbf{x}=(q,p,\mathbf{s})^T$ can be expressed as:
\begin{equation}
\!\left(\!\begin{array}{c}\dot{q} \\\dot{p}\\ \dot{\mathbf{s}}\end{array}\!\right)\!=
-\!\left(\!
\begin{array}{ccc}
0 & -1 & \mathbf	{0} \\
\omega_{0}^{2} & a_{pp} & \mathbf{a}_p^T \\ 
\mathbf{0} & \bar{\mathbf{a}}_p & \mathbf{A}
\end{array}\right)
\!\left(\!\begin{array}{c}q\\ p\\ \mathbf{s}\end{array}\!\right)\!+
\!\left(\!\begin{array}{ccc}
0 & 0& \mathbf{0}\\
0 & \multicolumn{2}{c}{\multirow{2}{*}{$\mathbf{B}_p$}}\\
\mathbf{0} & & \\
\end{array}\!\right)\!
\!\left(\!\begin{array}{c}0\\\multirow{2}{*}{$\boldsymbol{\xi}$} \\ \\\end{array}\!\right)\!
\label{eq:mark-harm}.
\end{equation}
Since the force is linear in $q$, equation \ref{eq:mark-harm} 
takes the form of an Ornstein-Uhlenbeck process,
that can be written concisely as
\begin{equation}
\begin{split}
\dot{\mathbf{x}} = -\mathbf{A}_{qp}\mathbf{x} + \mathbf{B}_{qp} \boldsymbol{\xi}.
\end{split}
\end{equation}
Since its finite-time propagator is known analytically~\cite{gard03book},
it is possible to compute any time correlation function in terms 
of the drift and diffusion matrices $\mathbf{A}_p$ and $\mathbf{B}_p$. For instance, the 
vibrational density of states can be computed exactly by taking the Fourier transform of the velocity-velocity correlation function, and reads:
\begin{equation}
\mathscr{C}_{pp}(\omega,\omega_0) = \frac{1}{[\mathbf{C}_{qp}(\omega_0)]_{pp}} \left[\frac{\mathbf{A}_{qp}(\omega_0)}{\mathbf{A}^2_{qp}(\omega_0) +\omega^2} \mathbf{C}_{qp}(\omega_0)\right]_{pp} \label{eq:vvac-gle},
\end{equation}
where the stationary covariance matrix can be obtained by solving
the Riccati equation
$\mathbf{A}_{qp}\mathbf{C}_{qp}+
\mathbf{C}_{qp}\mathbf{A}_{qp}^T=
\mathbf{B}_{qp}\mathbf{B}_{qp}^T$.

\begin{figure*}[hbtp]
\includegraphics[width=1.00\textwidth]{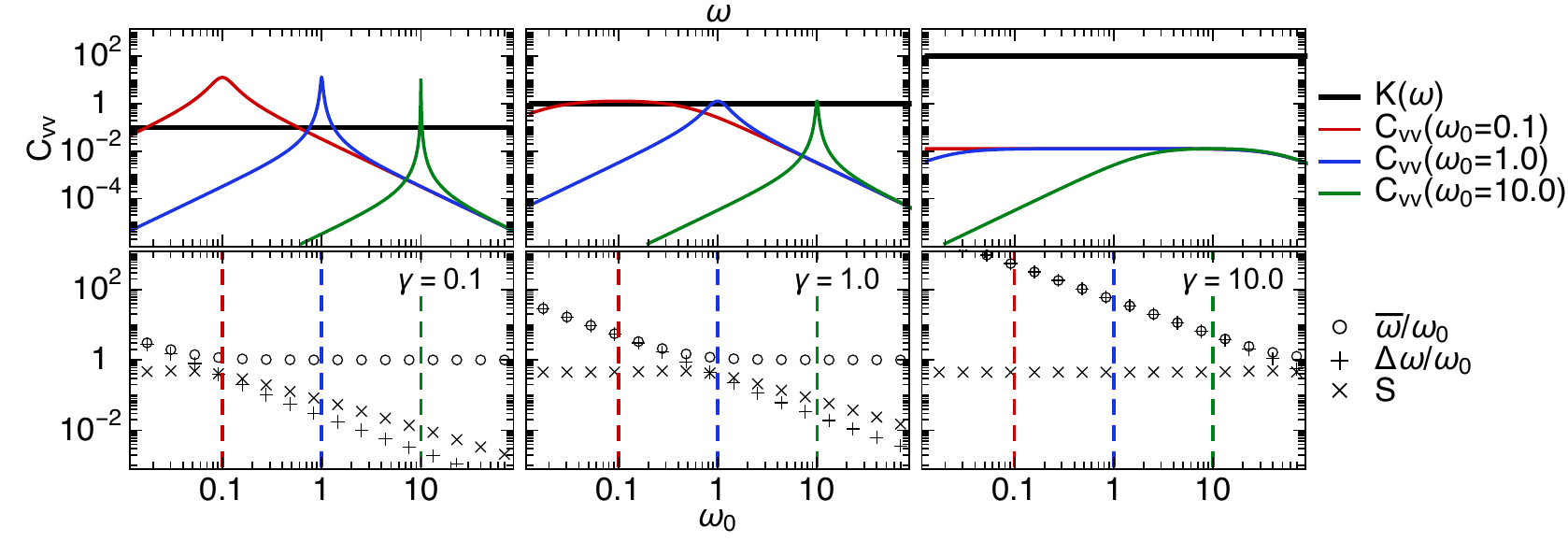}
\caption{Various regimes of the white noise thermostat acting on the harmonic oscillator. The left, central and right panels, respectively representing the under-damped ($\gamma=0.1$), optimally-damped ($\gamma=1$) and over-damped limits ($\gamma=10$), show the velocity auto-correlation functions (top)  and GLE metrics (bottom) for various values of the physical frequency. We choose three values for the physical mode $\omega_0$, labeled in the picture and shown with different colors. The GLE metrics as defined by equations \ref{eq:median}, \ref{eq:interq}  and \ref{eq:lor} are represented by circular, plus shaped and cross shaped markers respectively. 
\label{fig:harm-white-effect}}
\end{figure*}

It is useful to perform a spectral decomposition of Eq. \ref{eq:vvac-gle} in order to gain more insight into the spectrum a GLE-thermostatted oscillator. 
It is straightforward to show that by writing $\mathbf{A}_{qp}(\omega_0)=\mathbf{O}\operatorname{diag}(\mathbf{\Omega})\mathbf{O}^{-1}$ where $\mathbf{O}$ is the matrix of eigenvectors and $\mathbf{\Omega}$ a vector
containing the corresponding eigenvalues, the expression for the velocity-velocity correlation function can be written as
\begin{equation}
\mathscr{C}_{pp}(\omega,\omega_0) =  \sum_{jr}O_{pj}\frac{\Omega_j(\omega_0)}{\Omega_j^2(\omega_0) +\omega^2}  O^{-1}_{jr}\frac{[\mathbf{C}_{qp}(\omega_0)]_{rp}}{[\mathbf{C}_{qp}(\omega_0)]_{pp}} \label{eq:vvac-gle-decomp}.
\end{equation}

The spectrum in Eq. \ref{eq:vvac-gle-decomp}
corresponds to a sum of Lorentzian functions, with the peaks positions and lineshapes determined by the poles at $\omega=\pm i \Omega_j$.
Motivated by this spectral decomposition, we define several quantities that give a concise description
of the shape of the spectrum.
After having introduced the 
integral function of the spectrum
\begin{equation}
\begin{split}
W(\omega_a,\omega_b)&=\displaystyle{\frac{2}{\pi} \int_{\omega_a}^{\omega_b} \mathscr{C}_{pp}(\omega,\omega_0) d\omega =}  \\ &= \left\{ \left[ \tan^{-1}\left( \frac{\omega}{\mathbf{A}_{qp}(\omega_0)} \right)  \right]^{\omega_a}_{\omega_b} 
\frac{2 \mathbf{C}_{qp}(\omega_0)}{\pi [\mathbf{C}_{qp}(\omega_0)]_{pp}}
\right\}_{pp},
\end{split}\label{eq:cdf}
\end{equation}
which can be computed easily based on the same 
eigendecomposition of $\mathbf{A}_{qp}$,
we define the median
\begin{equation}
\bar{\omega}(\omega_0) \to W(0,\bar{\omega}) = 0.5, \label{eq:median}
\end{equation}
that characterizes the position of the peak, and the interquartile distance
\begin{eqnarray}
\Delta{\omega}(\omega_0)&=& \frac{1}{2}(\omega_{0.75} - \omega_{0.25}) \label{eq:interq} \\
&\to& W(0,\omega_{0.25}) = 0.25 \nonumber \\
&   & W(0,\omega_{0.75}) = 0.75 \nonumber
\end{eqnarray}
that characterizes its width. 
Together, these two indicators are sufficient to determine fully a Lorentzian lineshape
\begin{equation}
L(\omega,\omega_0)=\displaystyle{\frac{1}{\pi}\frac{\Delta \omega(\omega_0)}{(\omega-\bar{\omega}(\omega_0))^2+[\Delta\omega(\omega_0)]^2}}. \label{eq:lor}
\end{equation}
In order to quantify 
the presence of multiple 
poles or other sources of
asymmetry in the lineshape
that are not captured by 
$\bar{\omega}$ and $\Delta\omega$, 
we introduce a ``non-Lorentzian-shape"
factor $S$, 
\begin{equation}
S(\omega_0)=\left\vert \int_0^\infty [\mathscr{C}_{pp}(\omega,\omega_0) - L(\omega,\omega_0)]^2d\omega \right\vert^{0.5} \label{eq:nonlor}.
\end{equation}

According to the definitions above, a perfect $\delta$-like Lorentzian spectrum would have 
$\bar{\omega}/\omega_0=1$, $\Delta\omega/\omega_0=0$, and $\text{NL}=0$.
In order to exemplify how these measures behave in the case of a simple white noise
thermostat attached to the harmonic oscillator, we show in Fig. \ref{fig:harm-white-effect}
how the velocity-velocity spectrum of oscillators of different frequency $\omega_0$ changes with different regimes
of white noise, and how the measures defined in Eqs. \ref{eq:median} to \ref{eq:nonlor}
relate to the magnitude of the perturbation induced to a $\delta$-like spectrum shape.

Analyzing Fig. \ref{fig:harm-white-effect}, we can see that, as expected, the regime that introduces the least disturbance to the
VDOS is the underdamped regime (the limit where $\mathbf{A}_p=0$ is microcanonical dynamics) -- and that for a given $\gamma$ the modes with lower frequency suffer the most pronounced relative disturbance.
Focusing on the underdamped case, the measures $\bar{\omega}/\omega$ and $\Delta \omega$ predict the shift and broadening of the peaks at low frequencies, 
as well as the lack of disturbance at high frequencies. 
Going to the optimally damped and the overdamped case, the disturbances to the spectra
get more pronounced through the whole range of frequencies, and it is easy to follow
how the different
indicators we introduced quantify this change.
The $S$ measure is
always relatively small, indicating that 
a simple white-noise thermostat does not affect significantly the Lorentzian character of the peaks.

In the same spirit as the fitting procedure introduced in Ref. \cite{ceri+10jctc}, we
define figures of merit that target these measures, 
and complement the indicators of sampling efficiency that
were previously introduced.
By giving different weights to different targets and to
different frequency ranges, it is possible to generate GLE thermostats that are \emph{designed} to have a prescribed
effect when applied to a given system.
As we will show
below, even in cases for which the GLE thermostat disturbs
classical molecular dynamics in quite extreme ways, 
based on the analytical prediction of such disturbance one can recover the true dynamics of the underlying system.

\begin{figure*}[hbtp]
\includegraphics[width=\textwidth]{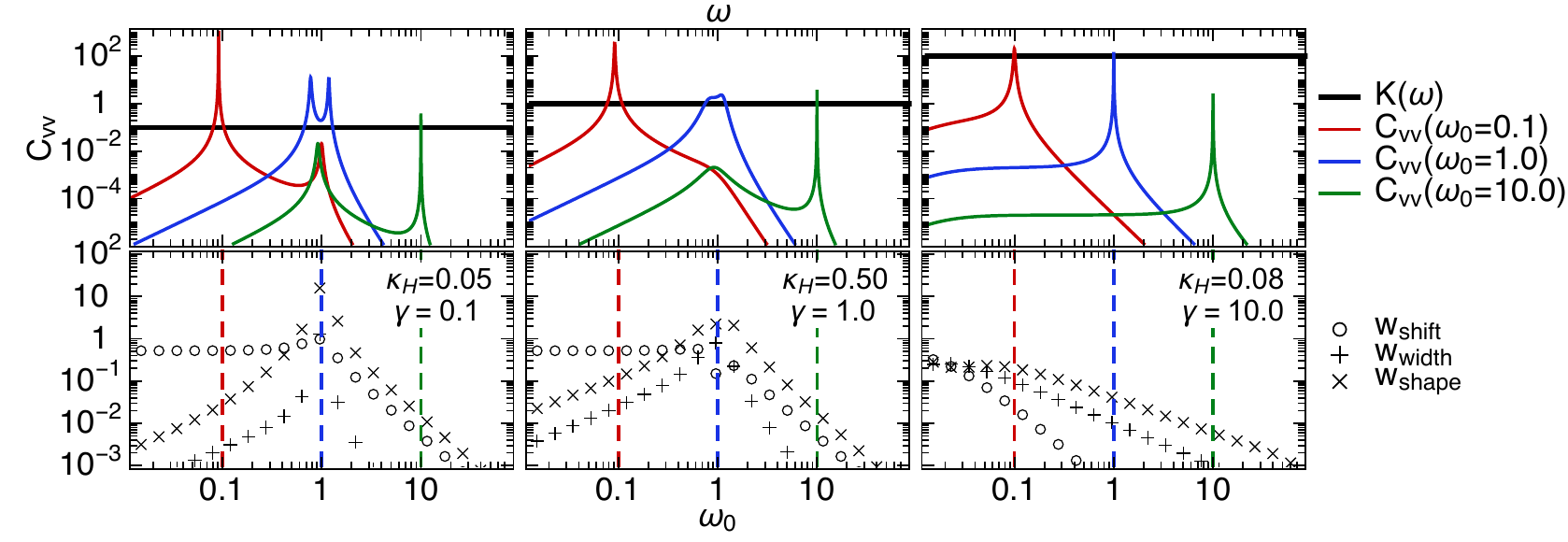}
\caption{Various regimes of the white noise thermostat applied to the ``ring-polymer'' mode. 
The left, central and right panels, respectively representing the under-damped ($\gamma=0.1$), optimally-damped ($\gamma=1$) and over-damped limits ($\gamma=10$), show the velocity auto-correlation function of physical mode (top) and GLE metrics (bottom) for various values of the physical frequency. We choose three values for the physical mode $\omega_0$, labeled in the picture and shown with different colors. The GLE metrics as defined by equations \ref{eq:rp-measure1}, \ref{eq:rp-measure2}  and \ref{eq:rp-measure3} are represented by circular, plus shaped and cross shaped markers respectively.
\label{fig:trpmd-white-effect}}
\end{figure*}

\subsection{Tuning Thermostated Ring Polymer Molecular Dynamics}

As shown in Ref. \cite{ross+14jcp}, the 
formalism underlying thermostatted ring polymer molecular dynamics (TRPMD)
leaves considerable
freedom into the way thermostats are applied to the 
internal modes of the ring polymer. In the original algorithm, a simple white
noise thermostat was used, that was tuned to give optimal sampling of the free ring polymer potential energy. 
Other choices for the white noise friction have been proposed, 
for instance attempting
to slow down the vibrations of internal modes to match those of the centroid\cite{hele+15mp}.
Here we show that
by optimizing quantitative 
measures of the interference
of ring-polymer modes onto
the dynamics of the centroid one 
we can improve the outcome of TRPMD simulations in a wide range of systems.

Let us start by introducing a simple model of the coupling of a ring polymer
mode to a physical (centroid) mode that we can use as the target of the GLE parameter optimization.
We examine the OU process of two coupled harmonic
oscillators where the $\mathbf{s}$ degrees of freedom are coupled to only one of them.
The potential thus has the form
\begin{equation}
V(q_0, q_1)= \frac{1}{2} \left[ \omega_{0}^2 q_0^2 + \omega_{1}^2 q_1^2  + \alpha \,\omega_0 \,\omega_{1} \,q_0 \,q_1  \right].
\end{equation}
From here on we denote $\omega_0$ the frequency of vibration of the physical system, $\omega_{1}$ the frequency of vibration of the ring polymer mode
that we wish to couple a thermostat to, and $\alpha$ a parameter that controls the strength of the coupling. 
Obviously one could redefine the physical coordinates to obtain two decoupled normal modes. Here instead we analyze the dynamics of the original coordinates, so that the harmonic coupling serves as an analytically-treatable model of anharmonic coupling.  

Extending the notation introduced in Eq.~\ref{eq:notation},
the drift matrix $\mathbf{A}_\text{01}$ for this system can be written as
\begin{equation}
\mathbf{A}_\text{01}=
\!\left(\!\begin{array}{cccccc}
 0  & -1 &  0  &  0  & 0 & \arS \\
 \omega_{0}^2 &  0  & \alpha \omega_0 \omega_1 & 0 &  0 & \arS \\
 0  &0  &  0  & -1 & 0 & \arS \\
 \alpha \omega_0 \omega_1   & 0 & \omega_1^2  & a_{pp} & \mathbf{a}_p^T & \arS \\
 0  & 0 & 0   & \bar{\mathbf{a}}_p &  \mathbf{A}  & \arS \\
\end{array}
\!\right)\!
\label{eq:notation-rpmd},
\end{equation}
where we maintain the same notation for the GLE drift matrix $\mathbf{A}_p$, with the understanding
that it only couples to $p_1$. In order to measure the disturbances on the physical system, we use 
indicators similar to the ones in Eqs. \ref{eq:median}--\ref{eq:nonlor}, but slightly
modified to capture the essence of this coupled-oscillators problem. 
Firstly, we can obtain analytical expressions for $\bar{\omega}$, $\Delta\omega$, and $S$ defined as in Eqs.\ref{eq:median}--\ref{eq:nonlor}, 
but referring to the power spectrum for the momentum
$p_0$ of the physical mode, $\mathscr{C}_{p_0p_0}$.
These three quantities depend parametrically
on $\mathbf{A}_p$, $\omega_1$ and $\alpha$.
In order to formulate the problem of optimising
$\mathbf{A}_p$ in a more general way, we first
consider that $\omega_1$ can be taken as the
reference frequency relative to which one considers
the frequency of the physical mode (i.e. we set
$\omega_1=1$ and aim to minimize the disturbance
for all $\omega_0$, smaller or larger than 1).
Our final optimized matrices can be easily scaled
by the target value of $\omega_1$ to which one wishes to 
attach the thermostat in a real calculation. In this work,
we scale the matrices by the  free ring polymer frequencies $\omega_k=2 \omega_P\sin(k\pi/P)$,
where $\omega_P=P/(\beta\hbar)$, $\beta=1/k_BT$, and $P$ is the number of beads in the ring polymer.

Since $\alpha$ is meant to represent a weak coupling term
we normalize our indicators based on their behavior
for small $\alpha$. This provides the following 
normalized target quantities:
\begin{eqnarray}
& w_\text{shift}& =(1-\bar{\omega}(\omega_{0};\omega_1)/\omega_0)/\alpha^2 \label{eq:rp-measure1}\\
& w_\text{width}& =\Delta{\omega}(\omega_0; \omega_1)/(\omega_0 \alpha^2) \label{eq:rp-measure2} \\
& w_\text{shape}& =S(\omega_0;\omega_1)/\alpha^2, \label{eq:rp-measure3}
\end{eqnarray}
\noindent that depend weakly on $\alpha$ (for numerical stability,
we took $\alpha=0.4$ in all of the calculations shown here).
As measured by this quantities, a ``perfect'', unperturbed
spectrum should yield $w_\text{shift}=w_\text{width}=w_\text{shape}=0$.

One particular issue that we wish to address with this procedure is the artificial
broadening of the peaks that is apparent in the original formulation
of TRPMD \cite{ross+14jcp}, that is especially bothersome
for the spectra of molecules. The origin of this broadening can 
be understood by analyzing, for example, 
how the simplified model described by
the drift matrix $\mathbf{A}_{01}$ behaves
when one uses a simple white noise thermostat
($a_{pp}=\gamma\neq0, n_s=0$).
In Figure \ref{fig:trpmd-white-effect} 
we show the quantities $w_\text{shift}$, $w_\text{width}$ and $w_\text{shape}$, as well as
the Fourier transform of the friction kernel $K(\omega)$, and the predicted $\mathscr{C}_{pp}$ for three different values of $\omega_0$. In all cases, we fix $\omega_1=1$, and use 
$\alpha=0.4$, that represents a fairly strong coupling, to exacerbate the effect.

In the underdamped regime (left-most panels of Fig. \ref{fig:trpmd-white-effect}), first focusing on the plotted  $\mathscr{C}_{pp}$, we observe that when $\omega_0\ll\omega_1$, the peak at $\omega_0$ becomes slightly red-shifted and a second low-intensity peak appears at $\omega_1$. When $\omega_0=\omega_1$ the peak is split, corresponding to the well-known RPMD resonance problem, that is well captured by this simplified coupling model. When $\omega_0\gg\omega_1$, the physical peak is sharp and there is essentially no shift, but there is a residual (weak) 
resonance at $\omega_1$. The measures
introduced in Eqs. \ref{eq:rp-measure1}--\ref{eq:rp-measure3} reflect this behavior: 
The indicator $w_\text{shift}$ predicts a larger disturbance for $\omega_{0}<\omega_1$ than for $\omega_{0}>\omega_1$, while $w_\text{width}$
and $w_\text{shape}$ predict little broadening and
non-Lorentzian lineshape in both the $\omega_{0}<\omega_1$ and $\omega_{0}>\omega_1$ limits.
 When $\omega_{0}\approx\omega_1$, the indicators correctly predict that the shift of the physical peak is not large (since the splitting is rather symmetric), but the width of the full (split) peak becomes larger. 
 The large value of $w_\text{shift}$ indicates the large deviation from a Lorentzian lineshape.
When the white noise frction on the ring-polymer mode is increased (moving to the right in Fig. \ref{fig:trpmd-white-effect}), we observe that all indicators predict better spectra.
However, when the peaks are in perfect resonance ($\omega_{0}=\omega_1$), 
optimal damping ($\gamma=\omega_1$) is not sufficient. Even though the peak is not split anymore, its shape is far from Lorentzian, and one observes considerable broadening -- compatible with the empirical observations for TRPMD. A certain broadening is also observed when $\omega_0\ll\omega_1$. 
Within this model, the over-damped regime gives us the best result regarding our disturbance measures, which is also reflected in the predicted vibrational spectra. 
In that regime, the largest disturbance is observed at $\omega \approx 0$ and the rest of the spectrum is clean. 
Note, however, that one cannot over-damp indefinitely.
The further one goes in the overdamped regime for the ring polymer mode, the less efficient that mode is sampled, as can be measured by $\kappa_H=\tau_H/\omega_1$, where $\tau_H$ is the autocorrelation time of the total energy for the ring-polymer mode. 
For this quantity, optimal sampling corresponds to $\kappa_H=0.5$.
A very aggressive damping can make the simulation much less efficient and non-ergodic -- a problem that can be mitigated by including sampling efficiency among the optimization targets. 

What we will show in the following is that by using a colored noise thermostat, one can get better results
than with only white noise, even though the trade-off between disturbance and sampling efficiency always
appears. In practice we obtain the colored noise matrices by optimizing an objective function that 
combines the newly-introduced indicators of dynamical
disturbance, computed over a broad range of  physical mode frequencies, with certain sampling efficiency requirements for the ring-polymer mode, in the same framework as introduced in Ref.~\citenum{ceri+10jctc}. We find, however, that 
the optimization has a pronounced tendency of finding 
local minima that nevertheless yield similar performances, as we discuss in more detail in in Section \ref{sec:result-disc}.

\section{Results and discussion \label{sec:result-disc}}

\begin{figure*}[hbtp]
 \includegraphics[width=0.9\textwidth]{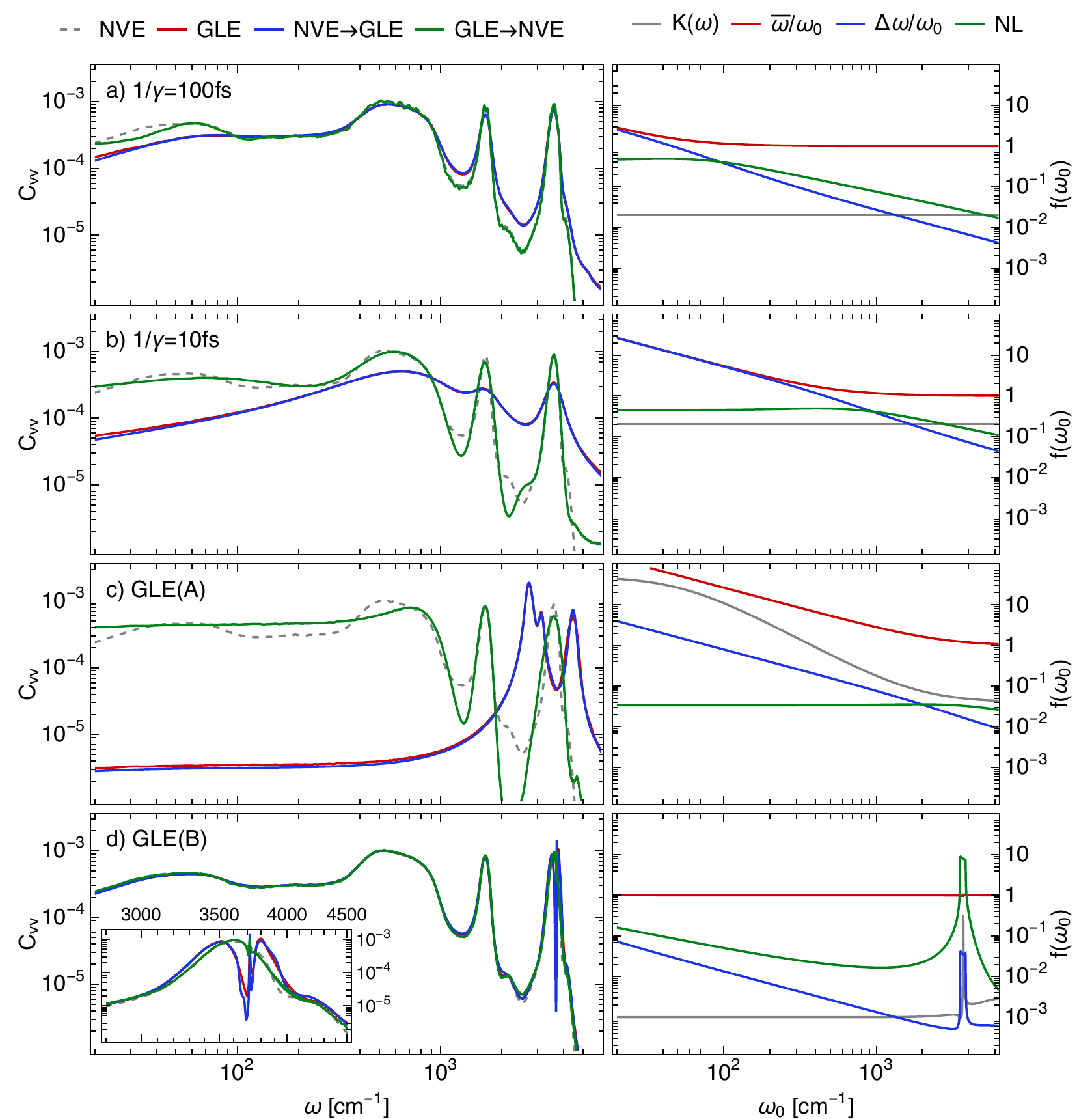}
\caption{
Each row reports the 
velocity-velocity correlation spectrum for a thermostatted simulation
of liquid water at 300K and experimental density (left) and 
the predicted measures of spectral disturbance ($\bar{\omega}/\omega_0$, $\Delta\omega/\omega_0$, $S(\omega)$) 
as a function of frequency, 
together with the GLE friction kernel $K(\omega)$ (right).  
The spectrum from the thermostatted trajectory (GLE) 
is compared with the density of states obtained from microcanonical runs (NVE), as well as with the spectrum predicted by convoluting the density of states with $\mathscr{C}(\omega,\omega_0)$ 
(NVE$\rightarrow$GLE) and the
density of states reconstructed
by deconvoluting the thermostatted spectrum (GLE$\rightarrow$NVE).
The simulations were performed with a strong white-noise thermostat (a),
a very-strong white-noise thermostat (b), a GLE designed to distort dramatically the whole spectrum (c), and a GLE designed to only affect the stretching peak (d with inset). 
\label{fig:water-nvt} }
\end{figure*}

In order to demonstrate the 
practical implications of the 
possibility of controlling the
impact of thermostatting on classical and quantum dynamics, 
we have computed the velocity-velocity correlation 
spectra of many different systems - including both gas-phase molecules and condensed phases of 
water. For the latter, we used 
a neural-network (NN) potential~\cite{behl-parr07prl,mora+16pnas} 
that has been fitted to reproduce 
a density-functional model of water~\cite{chen+16jpcl} based on the B3LYP  hybrid functional with D3 empirical dispersion corrections~\cite{grim+10jcp}, and 
that has been shown to reproduce accurately the first-principles results for many of the properties of liquid water~\cite{kapi+16jcp2}.
The general form of the NN potential ensures that the
anharmonicity of the ab initio
potential energy surface is 
fully reproduced -- making these
simulations a stringent test of the applicability of our analytical indicators beyond the harmonic limit.
All simulations presented in the following have been performed through
the interface of all relevant potentials with the i-PI code\cite{ceri+14cpc}.

\subsection{Predicting and correcting the dynamical disturbance of a GLE}

Equation~\ref{eq:vvac-gle} predicts 
the velocity-velocity correlation  
function for a harmonic oscillator of 
frequency $\omega_0$ subject to a given
GLE. If one considers an assembly of 
independent oscillators of different 
frequencies, the 
total correlation function of the 
system can be written as 
$\sum_i \mathscr{C}(\omega,\omega_i)$.
Taking the limit of a continuum
distribution corresponding to the 
density of states $g(\omega)$, one can 
write
\begin{equation}
c_{vv}^\text{GLE}(\omega)=\int \D\omega' g(\omega') \mathscr{C}(\omega,\omega'). \label{eq:cpp-convolution}
\end{equation}
Note that if rather than the total
velocity correlation function one were
computing a linear combination of correlation functions (e.g. a dipole spectrum to which each oscillator  contributes with its own transition dipole moment), 
Eq.~\eqref{eq:cpp-convolution} would 
still hold, 
with $g(\omega)$ representing a combination of the density of states and the weight of each mode. 
The question, of course, is how well
this relation would hold in a real, anharmonic system, and how well the 
indicators of dynamical disturbance
can be used to tune the behavior of the
GLE dynamics - given that the
kernel $\mathscr{C}(\omega,\omega')$
was derived under the assumption of 
harmonic dynamics. 
To benchmark this framework in 
a realistic scenario, we performed simulations of NN liquid water at 300K and experimental density.
We computed the vibrational density of states from a reference NVE simulation of the same model, and then compared 
it with the Fourier transform of the velocity-velocity correlation function 
resulting from different kinds of GLE.
Figure~\ref{fig:water-nvt} shows the 
results for white-noise Langevin 
dynamics using different values of the friction, and two GLE matrices (see the SI).
GLE(A) was designed to dramatically disturb all low-frequency modes,
whereas GLE(B) was optimized to only 
affect modes within a narrow range of 
frequencies between 3000 and 4000 cm$^{-1}$. 
Not only one can see that the GLE 
spectrum is qualitatively distorted
in accordance with the three indicators
$\bar{\omega}$, $\Delta\omega$ and $S(\omega)$, but also that convoluting the 
NVE density of states according 
to Eq.~\eqref{eq:cpp-convolution}
yields a near-perfect quantitative
prediction of the
GLE dynamics. 
These results open a path to the design
of thermostats that only affect a portion
of the frequencies while leaving the others 
untouched, as is the case for GLE(B).

Given the remarkable accuracy of the analytical prediction of the GLE dynamical disturbance, 
the possibility of performing the inverse operation arises --
that is to analytically 
predict the NVE density of states given
the velocity-velocity correlation function obtained from a thermostatted run. 
This operation corresponds to a 
deconvolution of the GLE spectrum using
$\mathscr{C}(\omega,\omega')$ as a 
convolution kernel. It is well-known
that this class of inverse problems 
is very unstable, and that an appropriate
regularization is crucial to obtain 
sensible results that are not dominated
by noise. 
Direct inversion using Tikhonov 
regularization with a Laplacian operator
led to promising but unsatisfactory results. In 
particular, we found a tendency to obtain large 
spurious oscillations in the low-density parts
of the spectrum, often leading to 
unphysical negative-valued curves.

\begin{figure*}[hbtp]
\includegraphics[width=0.9\textwidth]{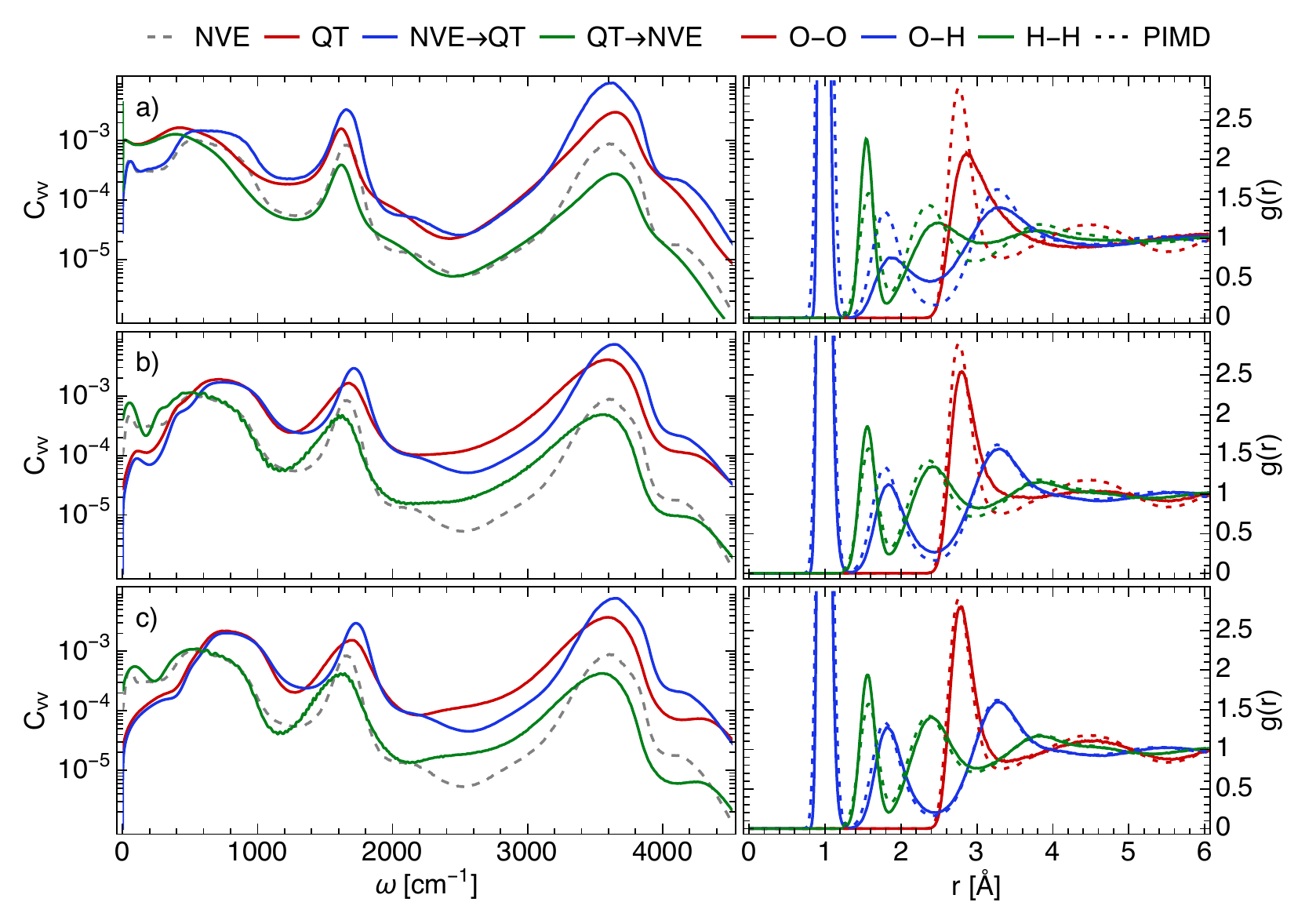}
\caption{ The panels
on the left report the velocity-velocity correlation functions, obtained from quantum-thermostatted simulations
of liquid water at 300K and constant experimental density.
As in Fig.~
\ref{fig:water-nvt}, the spectrum from a GLE simulation is compared with the NVE density of states, as well as with the transformed and reconstructed spectra. 
Panels on the right depict the radial O-O, H-H and O-H distribution functions from the QT runs, compared with those from a converged PIMD calculation \cite{kapi+16jcp2} (dashed lines). 
The topmost panels correspond to a weakly-coupled GLE, the middle and bottom panels correspond to 
strongly coupled GLEs fitted independently (see the SI for representative input files containing the parameters)
\label{fig:water-qt}}
\end{figure*}

We therefore used the Iterative Image Space 
Reconstruction Algorithm (ISRA), that enforces 
positive-definiteness of the 
solution\cite{daub-mueh86ieee,arch-titt95ss}. 
Initializing the iteration
with the GLE-computed velocity correlation spectrum, $f_0(\omega)=c_{vv}^{\text{GLE}}(\omega)$, 
the ISRA amounts at repeated application of the iteration
\begin{equation}
f_{n+1}(\omega)=
\frac{f_{n}(\omega)h(\omega)}
{ \int \D x \mathscr{D}(\omega,x) f_n(x) }
\end{equation}
where we have defined
\begin{equation}
\begin{split}
h(\omega)=&\int \D x \mathscr{C}(x,\omega) c_{vv}^\text{GLE}(x)\\
\mathscr{D}(\omega,x)=&
\int \D y \mathscr{C}(y,\omega)\mathscr{C}(y,x).
\end{split}
\end{equation}
The ISRA converges to a local solution satisfying 
$\int \D x\mathscr{C}(\omega,x) f_\infty(x)=c_{vv}^\text{GLE}(\omega)$.
We found that a convenient way to monitor the convergence
is to compute at each step
the residual, and the Laplacian of $f_n$,
\begin{equation}
\begin{split}
r_n&=\int\D\omega\left|\int \D x \mathscr{C}(\omega,x)f_n(x) 
-c_{vv}^\text{GLE}(\omega)\right|^2 \\
l_n&=\int\D\omega\left|f_n''(\omega)\right|^2.
\end{split}
\end{equation}

Plotting $(r_n,l_n)$ on a log-log 
scale reveals a behavior resembling 
a L-curve plot, that can be used as a 
guide to avoid over-fitting -- although
in practice we find that the 
well-known slow asymptotic convergence
of the ISRA effectively prevents
reaching a situation in which $f_n$ becomes 
too noisy. As can be seen from
Fig.~\ref{fig:water-nvt}, this approach
provides an excellent reconstruction
of the true density of states even in 
cases in which the GLE dynamics distorts the 
spectrum of water beyond recognition. 
There are of course discrepancies,  
particularly in the low-frequency region
that is both strongly anharmonic and 
harder to statistically converge. Nevertheless, the possibility
of correcting for the disturbance induced by a GLE
on the dynamics of complex atomistic system opens
up opportunities to obtain more accurate estimates
of dynamical properties from simulations that use 
Langevin equations to stabilize trajectories,~\cite{morr+11jcp} or that contain intrinsic stochastic terms ~\cite{kraj-parr06prb,kuhn+07prl,baer+13prl,mazz+14nc}.

\subsection{Dynamical properties from a quantum thermostat}

Besides correcting dynamical properties in classical thermostatted simulations, 
this iterative reconstruction of the unperturbed DOS
could be particularly helpful 
in another scenario. As mentioned in the 
Introduction, GLEs have been successfully applied
as a tool to sample a non-equilibrium 
distribution in which different vibrational modes
reach a stationary frequency-dependent 
effective temperature $T^\star(\omega)$. In particular, the so-called
``quantum thermostat''~\cite{ceri+09prl2} and 
``quantum thermal bath''~\cite{damm+09prl} try to 
enforce a temperature curve that mimics 
a quantum-mechanical distribution of energy in
the normal modes of the system. 
Trying to maintain this temperature imbalance in
an anharmonic system inevitably leads to zero-point 
energy leakage~\cite{habe-mano09jcp}, i.e. cross-talk
between different normal modes that lead to deviations
from the desired  $T^\star(\omega)$.
This problem can be addressed by using a
strongly-coupled GLE~\cite{ceri+10jctc}, that results however in 
a pronounced disturbance of the system's motion -- 
making any inference on quantum effects on dynamical
properties little more than guesswork. 
Being able to compensate for the dynamical disturbance
induced by a GLE can make this approach somewhat more
credible, and less dependent on the details of the
thermostat.

\begin{figure}[tbp]
\centering
\includegraphics[width=\columnwidth]{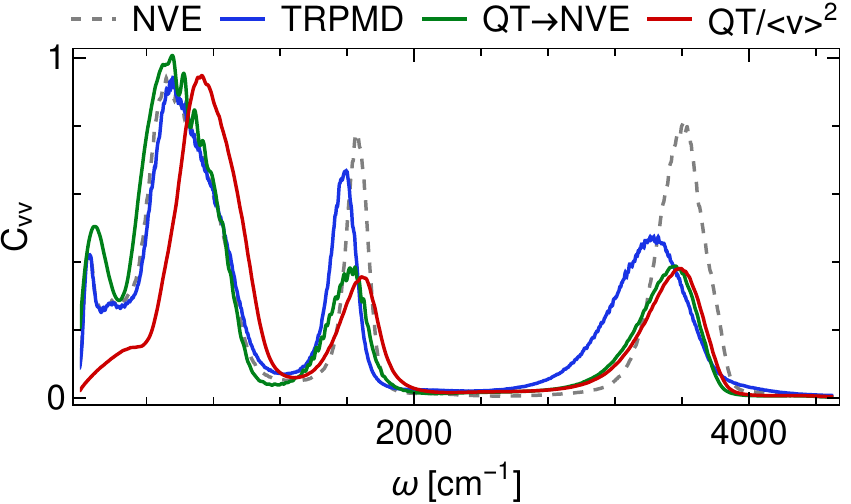}
\caption{A comparison between the classical vibrational density of states for a NN model of room-temperature water (NVE), 
that estimated from  critically-damped TRPMD (TRPMD), 
with the QT velocity-velocity correlation function scaled by $C_{pp}(\omega_0)$ (QT/$\left<v\right>^2$) and finally
the dynamically-corrected QT (QT$\rightarrow$NVE).
The QT parameters are those used for panel~(c) in Figure~\ref{fig:water-qt}.\label{fig:qt-trpmd}
}
\end{figure}

\begin{figure*}[btp]
\centering
\includegraphics[width=\textwidth]{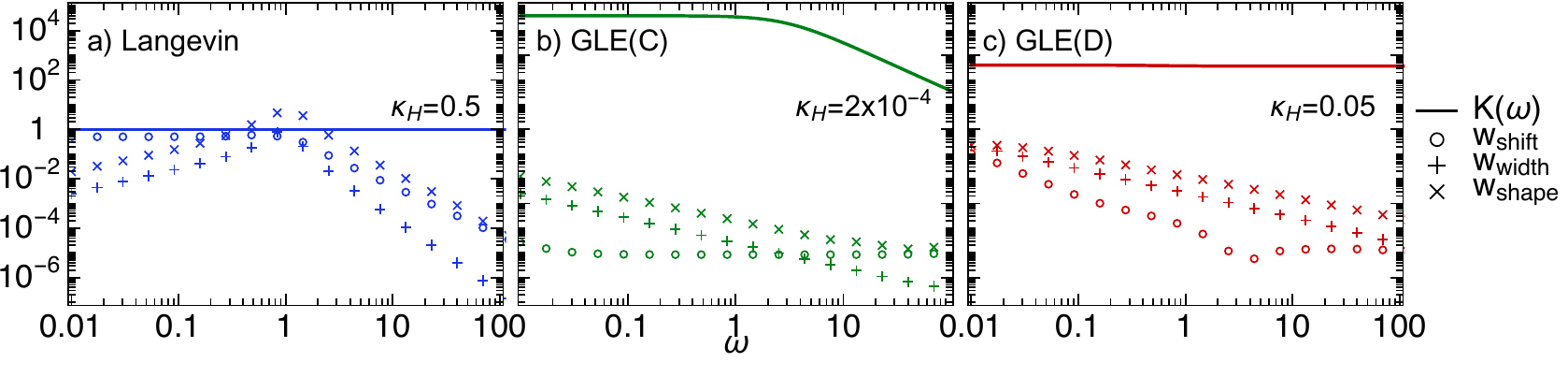}
\caption{Indicators as given in Eq. \ref{eq:rp-measure1}--\ref{eq:rp-measure3} for the GLE matrices used  to thermostat the internal modes of the ring polymers in the TRPMD simulations shown in this paper. \label{fig:gle-trpmd-matrices}}
\end{figure*}

Figure~\ref{fig:water-qt} gives a demonstration of
this idea -- as well as a clear warning to the 
dangers of using the results of a quantum GLE without careful 
validation. Let us start by discussing the accuracy of the QT in terms of structural properties, from which we can obtain a reliable benchmark from a fully converged~\cite{kapi+16jcp2} PIMD simulation of the same NN model. 
As seen from the radial distribution 
functions, using a weakly coupled quantum thermostat
(panel a) 
leads to significant zero-point energy leakage.
The stretching modes show narrower fluctuations
compared to PIMD, 
and the O-O distribution demonstrates a dramatic
loss of structure, which is compatible with a much
higher effective temperature of librational and translational modes. 
Increasing the 
coupling to the thermostat (panels b and c) improves significantly
the structure of water, that becomes very close to that from the PIMD simulation. This comes however at the price of 
a very pronounced 
disturbance of the dynamical properties, that is most 
apparent in the low-frequency part of $c_{vv}$.

Moving on to dynamical properties, let us now discuss the relations between the 
(classical) density of states, the GLE spectrum and
the curves obtained by convolution and deconvolution
through the kernel\footnote{It is useful to use a 
non-normalized kernel, as it automatically corrects for the different occupations of normal modes
of different frequency when converting between the density of states and the power spectrum.} $\mathscr{C}^\star(\omega,\omega_0)=m\beta C_{pp}(\omega_0)\mathscr{C}(\omega,\omega_0)$.
The deconvolution process corrects at the same time for
dynamical disturbances and the frequency-dependent occupations of different normal modes, so
any deviation between the reconstructed spectrum and the
classical DOS is an indication of anharmonic effects, and/or zero-point energy leakage that induces deviations
from the target $T^\star(\omega)$.
As shown in the lower panel of Fig.~\ref{fig:water-qt},
the iteratively-reconstructed DOS displays the
qualitative features one would expect from a quantum
spectrum of water: the low-frequency modes are effectively
unchanged relative to a classical DOS, whereas stretches
and bends show a considerable red shift and broadening. 
The reconstructed spectra from panels b and c -- that correspond
to different but strongly coupled GLEs -- are qualitatively very similar, particularly
when contrasted with the weakly-coupled GLE in panel a. 
In the 
latter case, the low-frequency modes are overheated, leading to an overestimation of the DOS relative to the classical limit, and the stretching
peak shows a blue shift, consistent with the fact that H-bonds are broken and stretch modes are underpopulated compared to the true quantum distribution.

\begin{figure*}[btp]
\includegraphics[width=\textwidth]{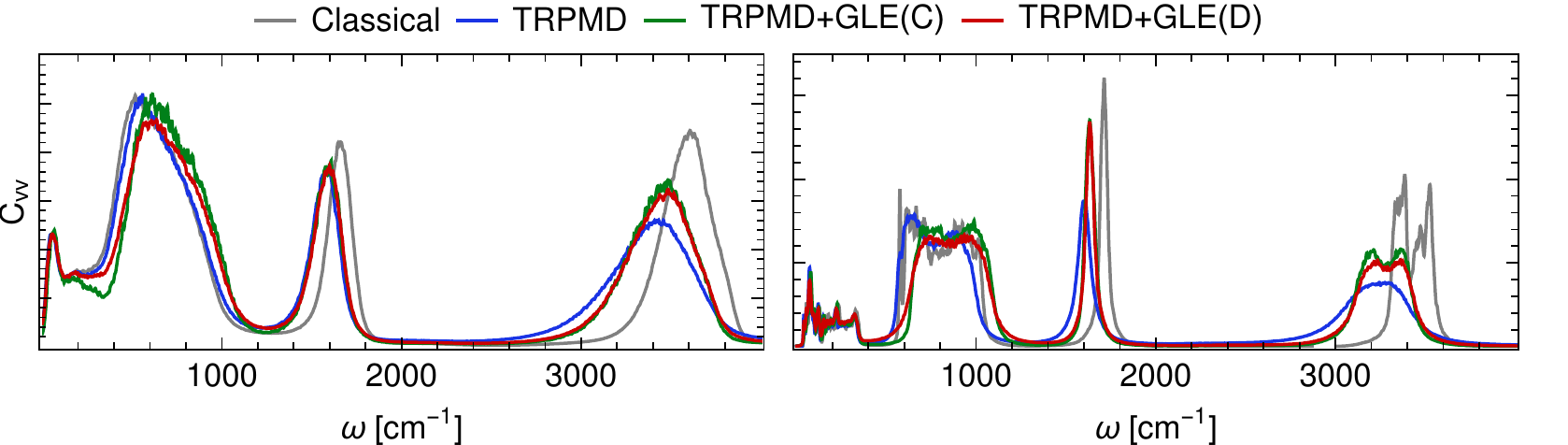}
\caption{Vibrational density of states for bulk water at 300 K (left) and and hexagonal ice at 100K (right) calculated with classical MD and different flavors of TRPMD using the neural network potential based on B3LYP+D3 reference data. \label{fig:gle-trpmd-condensedcvv}}
\end{figure*}

While there is no absolute benchmark for quantum effects on dynamical properties, it is useful to compare the results from the ``dynamically-corrected'' QT simulations
with those from a TRPMD simulation. 
As shown in Figure~\ref{fig:qt-trpmd}, the dynamical corrections do much more than rescaling frequencies by 
the QT occupations $C_{pp}(\omega_0)$. The heavily-distorted low-frequency part of the spectrum becomes very close to the classical 
DOS, and small corrections are also applied to  stretches and bending. 
While there is a considerable difference between the TRPMD spectrum and the corrected QT spectrum in the bending and stretching region, one should note that a similar discrepancy can be seen between TRPMD, CMD and other approximate quantum dynamical techniques~\cite{ross+14jcp2}. 
As we will show in Figure~\ref{fig:gle-trpmd-condensedcvv}, one can observe a similar degree of frequency shift when using a modified TRPMD designed to minimize dynamical artifacts.

We conclude this analysis by stressing that even though we showed examples based on the quantum thermostat,  
a similar analysis is possible for the case of a quantum thermal bath, which, even if implemented differently,  
is just a special case of the GLE framework in which the friction kernel is taken to be a $\delta$ distribution. 
Even though, whenever possible, one should cross-validate
results with a more sophisticated technique such as 
CMD or (T)RPMD, the dynamical corrections we introduce
to the quantum thermostat provide a practical 
solution for the cases in which one needs to 
assess the importance of quantum effects on 
dynamics but cannot afford a more accurate
method.

\subsection{Improving TRPMD spectra of molecular species}
 
As we discussed above, one can extend the GLE
model to assess the disturbance induced by the 
thermostatting of ring-polymer normal modes on
the dynamics of the centroid. We wish to assess
how the spurious broadening introduced by the white-noise
thermostat in TRPMD can be controlled and diminished using
GLE thermostats. We start by analyzing the vibrational density
of states of molecules, where the spurious broadening is particularly
dramatic.
We consider the isolated water molecule, simulated with the Partridge-Schwenke\cite{part-schw97jcp} 
potential, and the Zundel cation (H$_5$O$_2^+$), simulated with the CCSD(T)-parametrized potential of Ref. \cite{HuangBowman2005}. 
We performed all simulations at 100K, where nuclear quantum effects become more apparent, and used 64 beads to ensure convergence of the quantum distribution.

\begin{figure*}[ht]
\includegraphics[width=0.8\textwidth]{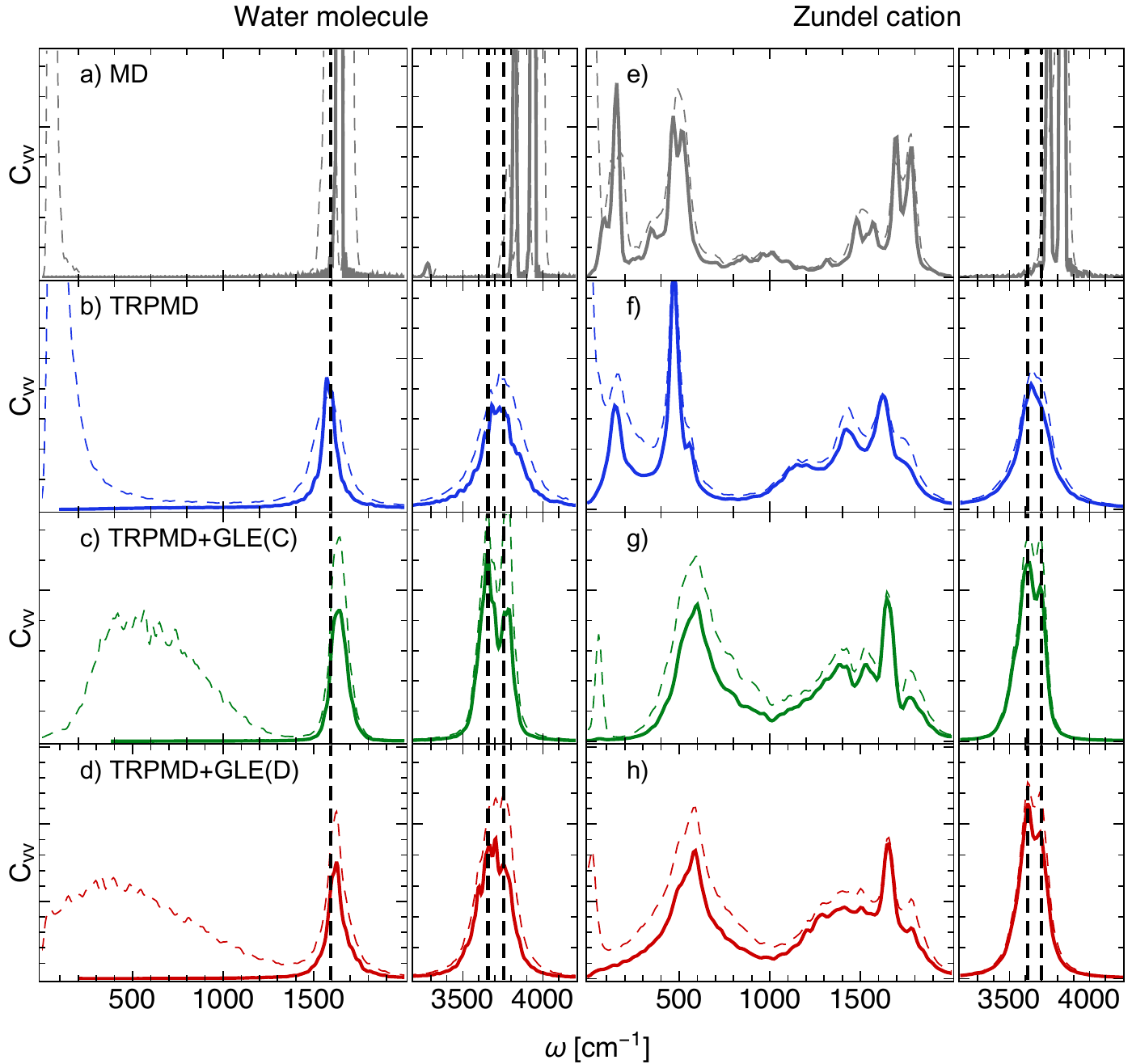}
\caption{Vibrational density of states at 100 K of the water molecule and the zundel cation calculated on the Partridge-Schwenke\cite{part-schw97jcp} potential and the potential of Bowman and coworkers\cite{HuangBowman2005}, respectively. Panels a to d show classical nuclei MD, white noise TRPMD, TRPMD+GLE(C) and TRPMD+GLE(D) vibrational density of states of the water molecule, and panels e to h show the respective vibrational density of states for the zundel cation. For each case we show in dashed lines the spectra including rotational motion, and in full lines spectra where these rotations have been filtered out. The reference data corresponds to the one reported in Ref. \cite{LiGuo2001} for the water molecule and in Ref. \cite{VendrellMeyer2007} for the Zundel cation.
\label{fig:trpmd-molecules}} 
\end{figure*}

Besides performing reference calculations with optimally-coupled white-noise, we tested the behavior of two GLE matrices, that were fitted
to minimize the analytical measures of dynamical disturbance for centroid modes with frequencies two orders of magnitude above and below the ring-polymer frequency. We also optimized the sampling of
the ring-polymer distribution as measured by the normalized autocorrelation rate $\kappa_H=1/\omega\tau_H$ of its total harmonic energy, in order to ensure it was not drastically inefficient.
Depending on the weights given to the different
target quantities, and on the starting parameters,
the optimization can converge to different (local) minima. Even restricting ourselves to matrices corresponding to a single additional degree of freedom, 
we observed that similarly good performances -- as measured by our analytical estimators -- could be achieved with two distinct
classes of $2\times 2$ $\mathbf{A}_p$ matrices.
The first kind of matrices
 had large off-diagonal components corresponding to an exponential-like kernel \cite{ceri10phd}, 
while those of the second kind are essentially dominated by their white noise component.
We show results for one matrix of each kind that we show below:

\begin{equation}
\text{GLE(C)}/\omega_1=
\!\left(\!\begin{array}{cc}
 1.0  & -241.4  \\
 244.8 &  2.9   \\
\end{array}
\!\right)\!,
\end{equation}

\begin{equation}
\text{GLE(D)}/\omega_1=
\!\left(\!\begin{array}{cc}
 182.4  & -3.7  \\
 2.8 &  0.6   \\
\end{array}
\!\right)\!.
\end{equation}

The indicators for
these matrices are shown in Fig. \ref{fig:gle-trpmd-matrices}. It is apparent that both matrices yield very good (i.e. very low) $w_\text{shift}$,
$w_\text{width}$, and $w_\text{shape}$ for a wide range of frequencies, with GLE(C) being slightly better overall -- at the expense of a lower $\kappa_H$.
Since the matrices were fitted assuming a 
unit frequency of the ring-polymer mode, optimum
parameters for each normal mode were obtained by
multiplying the chosen $\mathbf{A}_p$ matrix 
by the free ring polymer frequencies
at the relevant temperature.

In Fig. \ref{fig:trpmd-molecules} we show the vibrational
spectra of the isolated water molecule (panels a to d) and the zundel cation (panels e to h) calculated
with classical nuclei MD, with white noise TRPMD, with TRPMD+GLE(C) and with TRPMD+GLE(D) . 
We show in dashed lines spectra calculated from simulation
allowing rotation of the molecule and in full lines 
spectra where these rotations were filtered by changing the reference frame
at each time step in a post-processing procedure. For reference we also show the exact frequencies
of vibration in the water potential, calculated at 0K from Ref. \cite{LiGuo2001} and the multi-configurational time-dependent Hartree (MCTDH) OH stretch frequencies for the Zundel
cation taken from Ref. \cite{VendrellMeyer2007}. 

Focusing first on the spectra for the water molecule, we observe a 
considerable red shift of the OH stretch frequencies due to nuclear quantum effects, and all TRPMD simulations
can capture this shift. White noise TRPMD, however, is slightly blue-shifted with respect to the exact results, while the new GLE TRPMD are basically on top of the reference. The over-broadening of white noise TRPMD is also clear -- it cannot distinguish 
the splitting between symmetric and anti-symmetric stretches. The GLE thermostats make especially the OH stretch peaks narrower, and GLE(C) is even able to describe the splitting of the peak. For the OH-bend peak, we observe a red shift of 10 cm$^{-1}$ for TRPMD and a
blue shift of 20--30 cm$^{-1}$ for GLE(C) and GLE(D), with respect to the exact result. 
White-noise and GLE thermostatting
of the ring-polymer modes appear to have an impact on rotational dynamics,
that is only seen in the spectra that 
have not been cleaned from molecular rotations.

\begin{figure}[btp]
\centering
\includegraphics[width=\columnwidth]{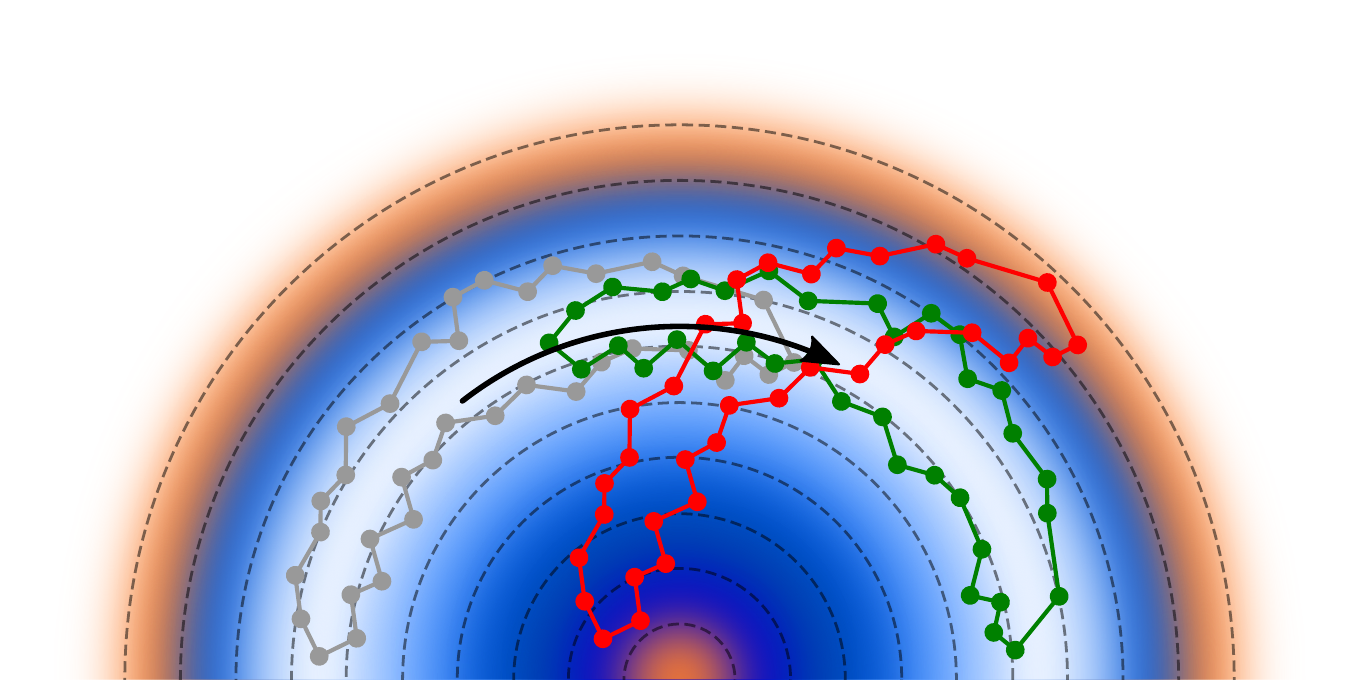}
\caption{
A cartoon representation of the origin of 
a blue shift of molecular degrees of freedom with a curvilinear nature, in presence of an overdamped dynamics of ring polymer modes. 
In order for the centroid to move along a curvilinear coordinate, the ring polymer (described in Cartesian coordinates) has to rearrange and change orientation (green polymer). 
If the internal motion of the ring polymer is hindered (red) the system tends to move rigidly, and it experiences a strong restoring force that increase the frequency of oscillation.\label{fig:champagne}
}
\end{figure}

Despite having been designed to minimally impact the dynamics of physical 
modes, the two GLE thermostats alter the spectral signature of rotations.
To qualitatively explain this effect, 
consider the cartoon representation of
the rotation of a ring polymer depicted in
Fig.~\ref{fig:champagne}. Similarly to what
is seen for the RPMD resonance problem and the CMD curvature problem, we here also introduce an artifact
that  is however associated with a curvilinear motion of the
centroid which is strongly coupled to the internal rearrangements of the ring polymer. 
In this case, the overdamped dynamics of the internal degrees of freedom of the path hinder the (near)-free rotation of the ring polymer, resulting in
an effective increase of the frequency of librations and rotations. 
A more quantitative analysis is far from
straightforward. The case of the rotational dynamics of a single particle subject to white noise is discussed in Ref.~
\citenum{WilkinsonPumir2011}. For a 2D rotor, one can compute analytically the orientational correlation function, that 
exhibits two qualitatively different 
regimes (Gaussian vs. exponential) in the limits where the friction $\gamma \to 0$ and $\gamma \to \infty$. 
The generalization to three dimensions 
is considerably more complex, and a 
formulation that considers coupling to a GLE (that would in
principle enable controlling and understanding these effects) is well beyond the scope of this work. Given that in most
practical applications quantum nuclear effects manifest themselves more strongly on
high-frequency modes that do not have a rotational character, this inconvenience
should not be particularly problematic.
We observe that increasing the importance of the optimization of $\kappa_H$ when designing the GLE does indeed
ameliorate this artifact: For example, the blue shift is less pronounced for GLE(D) (for which $\kappa_H=0.05$) than for GLE(C) ($\kappa_H=2\cdot 10^{-4}$). However, in our current optimization there is a trade-off between the sharpness of the spectra and the optimization of $\kappa_H$ (and thus the disturbance to the librations).

Moving now to the more complex spectrum of the Zundel cation 
we first focus on the high frequency range of the spectrum, where
nuclear quantum effects are expected to be most important. 
Comparing the classical, white noise TRPMD, and GLE TRPMD spectra, 
we observe that the GLE matrices behave according
to the desired specifications:
the peaks are sharper, making it possible to resolve the splitting between the OH stretch modes,  
and showing excellent agreement with the positions predicted by MCTDH for the same
potential energy surface. Focusing next on the low frequency range, 
we observe that the GLE thermostats cause a strong blue shift in the bands in that region, if compared to the classical and the
TRPMD simulations. These bands are in a region of the spectrum where nuclear quantum effects are expected to be very small, so
that the classical VDOS should be a good approximation in that region. The vibrations that populate the low
frequency range are related to librations of the water molecules in the complex with respect to each other and the central hydrogen.
Therefore, this is again a manifestation of the unphysical coupling of the GLE thermostats to overall curvilinear dynamics of the molecules.

\subsection{Asssessing the performance of TRPMD in the condensed phase}

Having analyzed the successes and shortcomings of GLE-thermostatted TRPMD simulations of molecules, 
we now assess their performance for condensed-phase simulations. The performance
of different quantum dynamics methods for the vibrational properties
of water at different state points has been assessed in Ref. \cite{ross+14jcp2} (where empirical potentials
were used), and the performance of path integral methods for the vibrational
properties of liquid water has been recently assessed on ab initio potential energy surfaces
in Ref. \cite{mars+17jpcl}. From these previous works, the conclusions were that
different types of quantum dynamics in the same potential could give results in good
overall agreement to each other and, regarding specifically TRPMD, that
also in the condensed phase it predicts high-frequency peaks that are considerably broader than predicted by other methods. It was also observed that quantitative details of the
impact of nuclear quantum effects on vibrational spectra depends strongly on the potential energy surface (something that
has also been noted for diffusion properties in water-based systems \cite{RossiMano2016} 
and for optical excitations \cite{SappatiGhosh2016}).

We use the same NN trained on the DFT-B3LYP+D3 potential energy surface 
that we used to compute classical and quantum-thermostatted spectra, and 
calculate the vibrational density of states of liquid water at 300K and ice Ih at 100K 
with TRPMD, using both optimally-damped white noise and the GLE
matrices discussed for the molecules. In Fig. \ref{fig:gle-trpmd-condensedcvv} we show these vibrational
spectra for water and ice (left and right
panels, respectively) and compare them with the classical density of states.

First, in this case the spectra of GLE(C) and GLE(D) are extremely similar for both liquid water and ice.
In more detail, starting from the OH stretch region, we observe a narrowing of the peak of
the vibrational spectra simulated with GLE(C)/(D) with respect to the white-noise TRPMD spectra.
The line
shape of the peak is also closer to the classical line shape. In the bend region, for
liquid water TRPMD and the GLE spectra agree almost perfectly, but for ice the GLE 
spectra predict much narrower peaks. For the libration band, 
we detect an unphysical blue-shift of the bands of both liquid water and
ice, with respect to the classical and TRPMD counterparts.
Similarly to what was observed for the molecules, the blue shift is (very) slightly more
pronounced for GLE(C), which induces a strongly overdamped dynamics on the ring-polymer vibrations.
Note that lattice vibrations of even lower frequency, as well as diffusion coefficients, do not suffer from these spurious effects. This confirms the that the curvilinear nature of the molecular motion plays a key role in these artifacts. 
Finally, we  observed that -- due to the relatively low sampling efficiency of the ring-polymer modes for GLE(C)/(D) -- it is somewhat harder to converge the populations of different normal modes 
when computing a vibrational density of states. 
In cases where this would constitute a problem, a simple solution consists in running multiple independent trajectories
off a single imaginary-time PIMD 
simulation~\cite{pere+09jcp}.

\section{Conclusions}

In this paper we have shown how Generalized Langevin 
Equation (GLE) thermostats can be used to manipulate 
the dynamical properties of physical systems in atomistic simulations,
not only when treating nuclei as classical particles, but also when modelling their quantum mechanical nature. 
We have introduced analytically-computable measures for the disturbance caused by the GLE to the
intrinsic dynamics of harmonic models. Based on these indicators, it is possible to obtain thermostats with very specific characteristics.

For molecular dynamics with classical nuclei,
where the GLE thermostats are coupled  directly to the physical system, 
we can calculate analytically the velocity-velocity correlation spectrum of a harmonic oscillator 
coupled to a GLE. We show that even in strongly anharmonic systems
such analytical predictions can be used to estimate the vibrational spectrum in the presence of the GLE,
through a convolution of the GLE spectrum with the underlying unperturbed
vibrational density of states. This observation also allowed us to deconvolute the velocity-velocity spectrum
from simulations run with a GLE thermostat, and recover the underlying unperturbed density of states. 
This deconvolution procedure
is particularly useful in all 
circumstances in which a degree of thermostatting is needed to stabilize the dynamics, or to compensate for 
random errors in the evaluated forces.
As an example, we consider the case 
of ``quantum thermostats'', that mimic
quantum statistical distributions by enforcing a frequency-dependent steady-state temperature on different normal modes. 
By correcting the dynamical disturbance introduced by the strong coupling of these thermostats (which is necessary to prevent zero energy
leakage), we put on more solid ground the practice of inferring dynamical information from these non-equilibrium, heavily thermostatted simulations.

When it comes to computing quantum dynamical correlation functions, 
one has to face the fact that no exact technique exists that can be taken as reference for condensed-phase 
(or large molecules) applications -- making it more difficult to determine objective measures of the quality of a thermostatted trajectory. 
Approximate techniques based on the path integral formalism\cite{cao-voth94jcp, crai-mano04jcp, ross+14jcp}
generally rely on performing classical dynamics for the ring-polymer centroid, on top of the quantum mechanical thermal distribution. 
The idea is then to guarantee that centroid dynamics are not affected by the behavior of the ring-polymer modes, that couple to the centroid by anharmonicities in the potential. 
We focused in particular on the thermostatted ring polymer molecular dynamics (TRPMD) method, 
since the underlying formalism leaves considerable freedom in choosing arbitrarily-complex thermostats to be attached to the internal degrees of freedom of the ring polymer. 
We designed an analytically-solvable model of the coupled centroid/internal mode dynamics, and computed estimators of the shift, broadening and general disturbance to the peak shape induced on the centroid. 
By optimizing these indicators for a broad range of centroid frequencies, we could significantly improve the quality of the vibrational spectra of gas-phase molecules -- in particular for the high-frequency portion that is most affected by quantum mechanical effects.

The GLE-optimized TRPMD density of states separates high-frequency peaks that were blurred in the white-noise version of the method,
and yields peak positions that correspond to the ones predicted by reference methods in the same potential energy surface. 
The vibrational density of states for 
condensed phases of water also shows sharper peaks in the stretch region.
We note, however, that our treatment introduced an unexpected (and unphysical) blue-shift on the rotational and librational modes
that seems unphysical, and that prevents us from making quantitative comments on this intermediate range of frequencies. 
We link this problem to the coupling between curvilinear motion of the 
centroid and the relaxation time of the internal modes of the ring polymer.
While in principle it might be possible to extend a GLE analysis to target rotational diffusion, and to reduce or remove this artifact, from a practical perspective the present GLE optimization is enough to improve the TRPMD spectra in the frequency range for which NQEs are most prominent. 
From a more fundamental point of view, it would be desirable to use the GLE framework in a less heuristic fashion, ideally deriving from first principles
the most appropriate form to approximate exact quantum dynamics.

In summary, we demonstrated that the very same GLE framework that has been successful for tuning the 
equilibrium sampling properties of classical and quantum molecular dynamics
can also be used to manipulate and correct the time-dependent behavior of a thermostatted trajectory. 
This approach can substantially extend the reach of many modelling approaches that rely on Langevin dynamics -- for instance, 
it is now possible to estimate diffusion coefficients, or vibrational spectra, from simulations performed in constant-temperature conditions. 
The fact we could also improve the quality of vibrational spectra obtained from approximate quantum dynamics techniques, 
based solely on the empirical goal of optimizing some measures of dynamical disturbance, 
underscores the potential of GLEs in this field. It also suggests that a more 
principled approach in deriving the appropriate form of the target memory kernels might inject additional physics into a family of methods that currently represent the 
most viable option to obtain time-dependent quantum mechanical observables for condensed-phase and large complex systems in general.

\section{Acknowledgements}

We thank David Manolopoulos for insightful discussion, and for comments on an early version of the manuscript. MR thanks 
a post-doctoral fellowship in the framework of the Otto Hahn Award of the Max
Planck Society for the fruitful time spent in Lausanne. VK and MC acknowledge the financial support by the Swiss
National Science Foundation (Project No. 200021-159896).

\section{Supplemental Information}

In the supplemental material we provide inputs for the i-PI program for all simulations presented in the paper. The inputs include the GLE matrices that we have optimized for each purpose. 

\bibliographystyle{aipnum4-1}
\end{document}